# Upper critical field in nanostructured Nb: Competing effects of the reduction in density of states and the mean free path

Sangita Bose, Pratap Raychaudhuri, Rajarshi Banerjee*and Pushan Ayyub

Department of Condensed Matter Physics and Material Science, Tata Institute of Fundamental Research, 1 Homi Bhabha Road, Mumbai 400005, India.

*Department of Materials Science and Engineering, University of North Texas, Denton, Texas 76203, USA

We show that the upper critical field in nanometer-sized Nb particles is governed by the changes in the effective Ginzburg-Landau coherence length occurring due to two competing factors: (i) the decrease in the grain size and consequent increase of disorder, and (ii) the effective decrease in the density of states at the Fermi level due to the formation of a Kubo gap. As a result, the upper critical field ($H_{C2}$) and irreversibility fields ($H_{irr}$) in nanostructured Nb show non-monotonic grain size dependences. Between 60nm to 20nm, $H_{C2}$ is found to increase by 2.5 times while there is no appreciable decrease in the superconducting transition temperature ($T_C$) from its bulk value of 9.4K. This can be ascribed to a decrease in the coherence length due to a reduction in the mean free path with decreasing size. Below 20 nm, however, $H_{C2}$ decreases with decreasing size. In this size range (<20 nm), there also occurs a decrease in the $T_C$ as well as the superconducting energy gap. The decrease in $H_{C2}$ in this regime can be ascribed to the decrease in the density of states at the Fermi level due to a quantization in the electronic energy levels.

## Introduction

Though granular superconductors have continued to draw attention over the past several decades, a complete and satisfactory understanding of the nature of the variation of superconductivity at reduced length scales is yet to be achieved. Among the numerous superconducting properties, the size dependence of the superconducting transition temperature ($T_C$) in particular, has been studied in great detail in a large number of systems including the elemental superconductors Al,[1] Sn,[2] Pb,[3] In[1] and Nb.[4] However, there are only a few studies of the size dependence of other macroscopic and microscopic superconducting properties such as the energy gap ($\Delta$), the London penetration depth, the coherence length and the upper and lower critical fields. Since some of these parameters of a superconductor are interrelated, a complete understanding of finite size effects can only be obtained when the evolution of each parameter is studied in detail.

As of now, there are only a few reports of the size dependence of the critical fields in superconductors,[5,6] and these are mostly on Type-I elemental superconductors such as Al and Pb. However, there are hardly any studies of the upper critical field ($H_{C2}$) in granular Type-II superconductors, though this is probably the property of utmost interest from the point of view of applications. The size-dependence of $H_{C2}$ in Type-II superconductors is particularly interesting since it is directly related to the density of states at the Fermi Level ($N(0)$) and the mean free path ($L_{eff}$), both of which are expected to decrease with decreasing particle size but have opposite effects on $H_{C2}$: a decrease in $N(0)$ should reduce $H_{C2}$, while a decrease in $L_{eff}$ should enhance it. It is therefore possible for $H_{C2}$ to show a non-monotonic dependence on particle size, but such a behavior has so far not been experimentally observed.

The Werthamer-Helfand-Hohenberg theory[7] relates $H_{C2}$ (T = 0) in the "dirty limit" ($L_{eff} \ll \xi_0$) to the normal state resistivity ($\rho_N$), the density of states at the Fermi level ($N(0)$) and the $T_C$ of the superconductor[7,8] in the form:

$$H_{C2}(0) = 0.69 T_C \frac{4 e c k_B}{\pi} N(0) \rho_N \qquad (1)$$

Here $L_{eff}$ is the mean free path, $\xi_0$ = intrinsic coherence length, while e = electronic charge and $k_B$ = Boltzmann constant. This expression suggests that increasing disorder (as might be caused by a decrease in the grain size) would lead to a larger $\rho_N$ and hence a monotonically increasing $H_{C2}$, provided $T_C$ and N(0) do not decrease appreciably with decreasing size. This formula, in fact, forms the basis of increasing the critical field in Type-II superconductors by the introduction of nonmagnetic impurities. A similar increase in $H_{C2}$ is expected in granular superconductors too, provided the $T_C$ does not change appreciably. Accordingly, there are reports of an enhancement in $H_{C2}$ of up to 60% in nanocrystalline Al[9] with a grain size of 4nm[10]. Recent studies on nanocrystalline $PbMo_6S_8$ report an increase in $H_{C2}(0)$ by almost 100% with a decrease in the grain size and hence in the mean free paths.[11] A similar enhancement in the low-temperature critical field has been observed in $K_3C_{60}$ powder and has been attributed to granularity as well as on intrinsic mechanisms such as strong electron-phonon coupling.[12] Note that in none of these systems is there is a size-induced decrease in $T_C$ in the relevant size regime. There are also a few reports of the enhancement of the critical fields in ultrathin films.[13,14,15] However, there appears to have been no systematic study of the size dependence of the critical field in nanocrystalline Nb, the elemental superconductor with the highest known $T_C$ (9.2K). Nb is known to show an appreciable decrease in $T_C$ with decreasing grain size below 20nm.[4] It is therefore important to investigate how the opposing influences of decreasing $T_C$ and increasing $\rho_N$ (due to reduction in grain size) affect the critical field in nanocrystalline Nb.

In this paper, we report the variation of the upper critical field as well as the irreversibility field in nanocrystalline thin films of Nb as a function of particle size. We observe a 2.5 times increase in the value of $H_{C2}$ at low temperatures, as the grain size is reduced from 60nm to 20nm. However, the size dependence of $H_{C2}$ is non-monotonic and $H_{C2}$ *decreases* with decreasing size below ≈20nm. We point out that this is the size range in which $T_C$ also decreases with size. $H_{irr}$ exhibits a non-monotonic grain size dependence similar to $H_{C2}$. In an earlier communication,[4] we had shown that the experimentally determined superconducting energy gap ($\Delta(0)$) in nano-Nb decreases below 20nm and scales with the $T_C$, indicating that N(0) too decreases with decreasing particle size. Thus, the size dependence of $H_{C2}$ results from two competing factors: a decrease in the mean free path and a destruction of the superconducting order parameter as the size of the superconductor is decreased.

## Experimental Details

Nanocrystalline Nb thin films (about 0.5μm thick) were deposited on Si substrates with an amorphous oxide overlayer, using high pressure magnetron sputtering.[16] Sputter deposition was carried out from a 99.99% pure Nb target in a custom-built UHV chamber with a base pressure ~$3\times10^{-8}$ torr. The average grain size in the nanocrystalline films was controlled by varying the Ar pressure in the 3-100 mtorr range and the dc power in the 25-200W range. The microstructure of the as-deposited Nb films was characterized by x-ray diffraction (XRD) and transmission electron microscopy (TEM). TEM studies were carried out in a FEI/Philips CM200 TEM and a FEI Tecnai F20 FEG-TEM, both operating at 200 kV. The strain corrected coherently diffracting domain size ($d_{XRD}$) was calculated from x-ray line profile analysis using the XFIT software. By controlling the deposition conditions, $d_{XRD}$ could be varied from 60 nm down to 5 nm. For different samples, the average grain size observed under bright or dark field TEM matches quite well with $d_{XRD}$. In this paper, all references to grain or particle size actually implies $d_{XRD}$. The samples being studied here consist of a dense aggregate of *bcc* Nb nanoparticles that can be visualized as a network of weakly connected superconducting grains. The upper limit of the particle size distribution obtained by the Warren-Averbach method (x-ray line profile analysis) was found to be no more than 20% of the mean size for any of the samples. Detailed structural characterization of the nanocrystalline Nb films by XRD and TEM have been described elsewhere.[17]

The upper critical field was obtained from transport measurements carried out in a custom built system with a variable temperature insert in the temperature range 1.6 – 300K. The standard four-probe method was used with the sample placed parallel as well as perpendicular to the magnetic field direction. The field ($H$) dependence of the resistance ($R$) was recorded at a series of different temperatures below $T_C$. We define the upper critical field ($H_{C2}$) as that value of $H$ at which the resistance falls to 90% of its normal state value.

The irreversibility field ($H_{irr}$) was estimated from both high frequency ac susceptibility ($\chi$) as well as dc transport measurements. The ac susceptibility set up consisted of two planar coils (acting as the primary and secondary, respectively) between which the sample was sandwiched. When the sample becomes superconducting, it shields the magnetic field produced by the primary from the secondary, and causes a sharp drop in the real part of the signal picked up by the secondary coil. This drop in signal is used to accurately determine the $T_C$, $H_{C2}$ and $H_{irr}$. While this is an extremely sensitive technique for determining these quantities, one cannot obtain the *absolute value* of the susceptibility from this configuration. The measurement was carried out at 15 kHz with an applied ac field of less than 1 Oe. An external superconducting magnet was used to superpose a dc field for the isothermal $\chi-H$ runs. We define the irreversibility field ($H_{irr}$) as that value of $H$ at which the resistance falls to 10% of its normal state value. It is known that the temperature at which the dissipation peak in the ac susceptibility ($\chi''$) occurs can be related to the temperature corresponding to the 10% normal state resistance obtained from magnetoresistive measurements.[18] For an accurate determination of $H_{irr}$ at temperatures close to $T_C$ (where the values of $H_{irr}$ are low), a dual magnet power supply was used to correct for the remanence of the superconducting magnet.

## Results and Discussions

The size dependence of the superconducting transition in nanocrystalline Nb has been discussed in detail in Ref. 4. Here we discuss aspects related to the upper critical field of the same system. Figure 1 shows representative magnetoresistance data (*R* vs. *H*) at different temperatures for the samples with grain size of 28nm and 11nm, recorded with the film placed parallel to the external field. A representative plot of the real ($\chi'$) and imaginary ($\chi''$) parts of the susceptibility of the nano-Nb film with a grain size of 17nm at a temperature of 2K is shown in Fig. 2. The *H-T* phase diagram was drawn for all the samples after estimating $H_{irr}$ and $H_{C2}$ at different temperatures using the method described in the previous section. Figure 3(a-d) shows such phase diagrams for four nano-Nb samples with grain sizes of 45nm, 19nm, 17nm and 11nm respectively. The irreversibility line (obtained from both magnetoresistance and ac susceptibility data) has also been plotted for the same samples in parallel and perpendicular orientations. As expected for conventional, low-$T_C$ superconductors the irreversibility line lies quite close to the $H_{C2}$-$T$ curve. It is also clear that the values of $H_{irr}$ determined from ac susceptibility and magnetoresistance match quite closely. Further, the values of $H_{irr}$ and $H_{C2}$ change very little with the orientation of the film with respect to the external magnetic field. We should not, in fact, expect any orientation dependence of the critical fields, since the nanocrystalline films are about 0.5μm thick and hence essentially behave as bulk systems with insignificant orientation dependence.

We observe that the critical field varies linearly with temperature for the samples with grain sizes < 20nm, as expected for granular superconductors. A detailed microstructural characterization by TEM and transport measurements had earlier shown that the nano-Nb films in this size regime behave as granular superconductors with the crystalline grains separated by relatively narrow intergranular regions consisting of an insulating Nb-O phase.[16] Hence, the films form a network of weakly-connected Josephson junctions. For Type-II superconductors, $H_{C2}$ is related to $\xi_{GL}$ as:[19]

$$H_{C2}(T) = \frac{\phi_0}{2\pi[\xi_{GL}(T)]^2} \qquad (2)$$

where $\phi_0$ is the flux quantum. Now, in the dirty limit ($L_{eff} \ll \xi_0$), where $\xi_0 = \frac{0.18\hbar V_F}{k_B T_c}$ ($V_F$ is the Fermi velocity) is the intrinsic coherence length, $\xi_{GL}(0)$ has the following temperature dependence near $T = T_C$:

$$\xi_{GL}(T) = 0.85 \frac{(\xi_0 L_{eff})^{1/2}}{(1 - T/T_C)^{1/2}} \qquad (3)$$

For granular superconductors in the dirty limit, $H_{C2}$ would therefore vary linearly with temperature near $T_C$.

Figure 4 shows the observed grain size dependence of $H_{C2}$ in nano-Nb. $H_{C2}$ increases from its bulk value of 2.8T for the film with the largest grain size (~60nm) that shows the bulk $T_C$ (9.4K) to 6.9T for the film with a grain size ~19nm with $T_C$ = 7.8K. However, for the films with grain size below 19nm, $H_{C2}$ decreases till it reaches a value of 5T for the smallest sized superconducting sample ($d_{XRD}$ = 11nm). Clearly, $H_{C2}$ exhibits a non-monotonic size dependence in nanostructured Nb.

The grain size dependence of $\xi_{GL}(0)$ was obtained by substituting the observed values of $H_{C2}$ in Eqn. 2. Expectedly, $\xi_{GL}(0)$ also shows a non-monotonic dependence with grain size. As a consistency check we made an independent determination of $\xi_{GL}(0)$ using the following relation obtained from the GL theory:[20]

$$\xi_{GL}(0) = \left[\frac{\phi_0}{2\pi T_C (dH_{C2}/dT)_{T_C}}\right]^{1/2} \quad (4)$$

Figure 5 shows the variation of $\xi_{GL}(0)$ with grain size obtained from both the above methods and it is clear that they agree within experimental error. Table 1 shows the coherence length ($\xi_{GL}(0)$) for the six samples with different grain sizes ($d_{XRD}$) and $T_C$. Note that the Ginzburg-Landau coherence length for the largest grain sample (60nm) still shows a much reduced value (see Table 1) than that of bulk Nb (41±3nm). This is because our 'bulk' samples are also in the dirty limit with very low residual resistivity ratio ($\rho_{300K}/\rho_{10K}$), which corresponds to mean free paths of the order of a few nanometers.

The increase in $H_{C2}$ by $\approx$ 2.5 times in the intermediate size regime (from 60nm down to 20 nm) can therefore be correlated with the size-dependent decrease in $\xi_{GL}(0)$. It is expected that with reducing grain size and increased grain boundary scattering the mean free path ($L_{eff}$) should monotonically decrease and hence also the coherence length. However this factor alone should lead to a monotonic enhancement of $H_{C2}$ with decreasing grain size, which is contrary to our observations. Thus, our results emphasize the fact that the coherence length is not determined by the grain size (or mean free path) alone, but also depends on the density of states, particularly in the quantum size regime.

From independent measurements of the $T_C$ and the superconducting energy gap ($\Delta(0)$) in the lower size regime (<20nm), we have earlier shown[4] that both $T_C$ and $\Delta(0)$ decrease *proportionately* with grain size. Thus Nb remains in the weak coupling limit with $2\Delta(0)/k_BT_C \approx 3.6$, down to the smallest particle size of 11nm. The decrease in $T_C$ for sizes <20nm (see inset of Fig. 4) can therefore be attributed to a change in the electronic density of states near the Fermi energy due to a size-induced discretization of the energy levels (Kubo gap). Note that quantum size effects may cause a decrease in the $T_C$ even for sizes (10-20nm in this case) for which the Kubo gap is still smaller than the superconducting energy gap. This is due to the decrease in the effective density of states at the Fermi level, $N(0)$. The Fermi level lies in between the last filled and the first unfilled electronic states. Therefore, when the spacing between adjacent levels is larger than the typical broadening of each level, there will be a decrease in the effective density of states at the Fermi level. It is clear from Eqn. 1 that a decrease in $N(0)$ at sufficiently small sizes would give rise to a concomitant decrease in $H_{C2}$, overriding the effect of decreasing mean free path. This results in the observed non-monotonic behavior of $H_{C2}$.

We point out that the monotonic increase in $H_{C2}$ observed in earlier studies of granular superconductors were on systems such as Al and Pb, which either show an increase in $T_C$ (in Al[9]) or no change (in Pb[21]). Hence, in such systems, the reduction in the mean free path with decreasing grain size plays the most dominant role in controlling the critical field. Interestingly, even a non-monotonic behavior of $H_{C2}$ has been observed earlier – not in nanocrystals – but in ultrathin films (< 10nm thick) of Nb.[12] Both $H_{C2\parallel}$ and $H_{C2\perp}$ (measured in fields parallel and perpendicular to the film plane, respectively) showed a non-monotonic dependence on film thickness. As might be expected in the light of the present study, those Nb films also exhibited a reduction in $T_C$ with film thickness.

Another system which shows a similar non-monotonic behavior of $H_{C2}$ however does not belong to the class of low dimensional superconductors. This is the Al doped $MgB_2$ system (both in bulk and thin film form)[22,23] where the $H_{C2}$ initially increases with Al doping and then decreases at higher doping levels. Al-doping is equivalent to an increase in the disorder in the system and leads to a reduction in the $T_C$. In such systems too, one observes a decrease in the superconducting energy gap and hence the effect is attributed to a decrease in the superconducting order parameter. It is useful to point out that MgB$_2$ is a multiple band superconductor where interband scattering plays a vital role in determining the properties. Hence, it should not be directly related to Nb, which is a single band conventional s-wave superconductor.

In conclusion, we have observed a large ($\approx$2.5 times) increase in the upper critical field in nanostructured Nb films as the grain size is reduced from 60nm to 20nm. Below 20nm, however, there is a decrease in H$_{C2}$. There is a similar non-monotonic decrease in the coherence length arising from a reduction in the mean free paths of the films with decreasing grain size. This anomalous and non-monotonic behavior of H$_{C2}$ has been related to the competition between the decrease in coherence length which tends to increase H$_{C2}$ and the decrease in the superconducting energy gap which tries to destabilize superconductivity when the grain size is reduced below $\approx$20nm. Our results thus provide a clear demonstration of the interplay of different superconducting parameters on the upper critical field of superconductors at reduced dimensions. The understanding of such competing interactions is critical if one attempts to use grain size as a parameter in optimizing the upper critical field for magnetic applications.

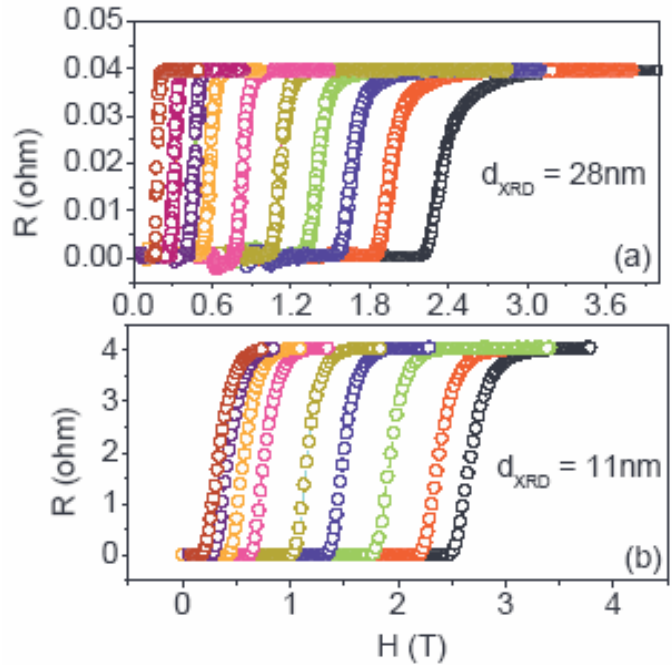

**Figure 1.** Magnetoresistance data for nanocrystalline Nb films with (a) $d_{XRD}$ = 28nm (b) $d_{XRD}$ = 11nm. Different curves refer to R vs. H data recorded at different temperatures. In (a), the temperatures (from right to left) are: 4.4, 5.0, 5.5, 6.0, 6.5, 7.0, 7.5, 8.0, 8.5, 9.0K respectively, while in (b) these values are: 2.6, 3.0, 3.5, 4.1, 4.6, 5.0, 5.3, 5.5, 5.6K respectively.

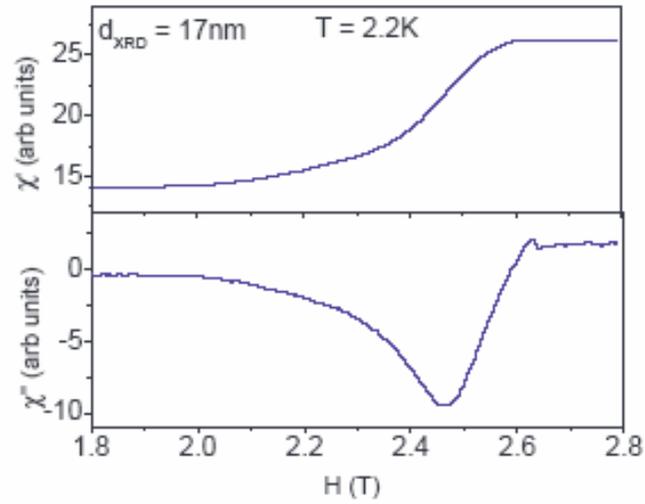

**Figure 2.** Real (top) and imaginary (bottom) parts of the ac susceptibility (χ) for a representative nanocrystalline Nb sample with $d_{XRD}$ = 17nm, recorded at $T$ = 2.2K

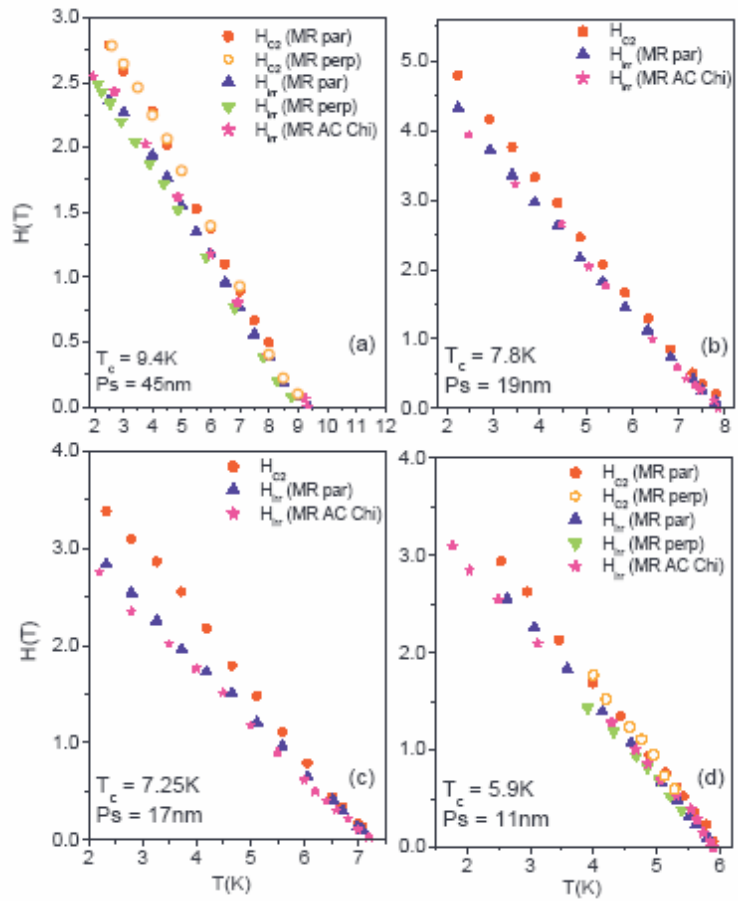

**Figure 3.** Variation of $H_{C2}$ and $H_{irr}$ with temperature (T) for the four nano-Nb films with (a) $d_{XRD}$ = 45nm, (b) $d_{XRD}$ = 19nm, (c) $d_{XRD}$ = 17nm, and (d) $d_{XRD}$ = 11nm respectively. The solid and open circles denote $H_{C2}$ with the samples in parallel and perpendicular orientation, upright and inverted triangles denote $H_{irr}$ in parallel and perpendicular orientations, respectively, all determined from transport measurements. The stars denote $H_{irr}$ obtained from ac susceptibility measurements.

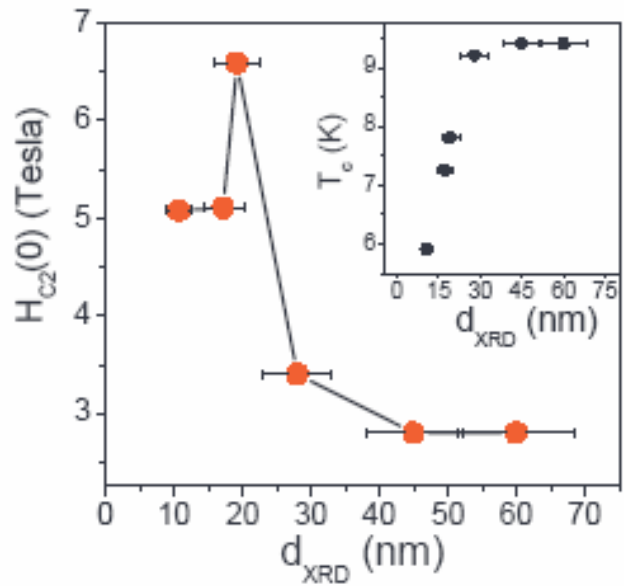

**Figure 4.** Variation of $H_{C2}$ with grain size ($d_{XRD}$) for the nano-Nb films. Note the break in monotonicity around 20nm. Inset shows the variation of $T_C$ with $d_{XRD}$ (from Ref. 4).

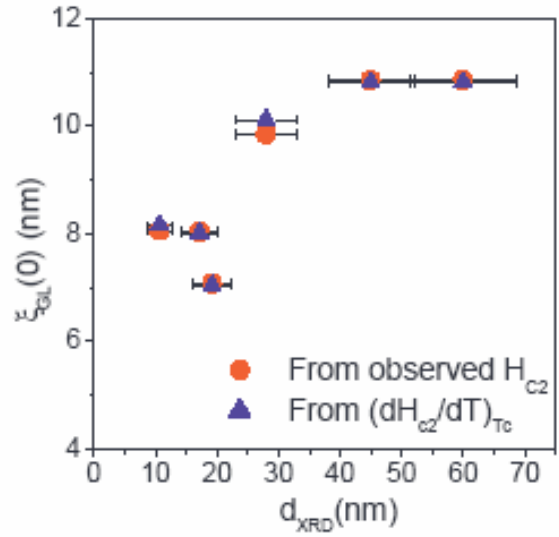

**Figure 5.** Grain size dependence of the Ginzburg-Landau coherence length, $\xi_{GL}(0)$ calculated from $H_{C2}$ (using eqn. 2) and also measured from the slope of the $H_{C2}$-T curve (using eqn. 4).

.

**Table 1.** Grain size dependence of various superconducting and normal-state properties of nanocrystalline Nb.

| $d_{XRD}$ (nm) | $T_C$ (K) | RRR $\frac{\rho(300K)}{\rho(10K)}$ | $H_{C2}(0)$ (T) | $(dH_{C2}/dT)_{Tc}$ (T/K) | $\xi_{GL}(0)$ (nm) |
|---|---|---|---|---|---|
| 60 | 9.4 | 8.4 | 2.8 | 0.3 | 10.8 |
| 45 | 9.4 | 6.4 | 2.8 | 0.3 | 10.8 |
| 28 | 9.2 | 4.2 | 3.4 | 0.35 | 9.8 |
| 19 | 7.8 | 2.3 | 6.6 | 0.85 | 7.03 |
| 17 | 7.2 | 1.88 | 5.1 | 0.70 | 8.03 |
| 11 | 5.9 | 1.49 | 5.1 | 0.84 | 8.06 |